\documentclass[twocolumn]{aastex631}

\received{Sep 01, 2022}
\revised{Sep 27, 2022}
\accepted{Nov 05, 2022}

\submitjournal{ApJ}
\shorttitle{Rotational \textcolor{black}{outflow}}
\shortauthors{Ko et al.}

\usepackage{physics}
\usepackage{bm}
\begin{document}

\title{\textcolor{black}{Self-Similar solution of rotating eruptive outflows on its equatorial plane}}

\correspondingauthor{Takatoshi Ko}
\email{ko-takatoshi@resceu.s.u-tokyo.ac.jp}

\author{Takatoshi Ko}
\affiliation{Research Center for the Early Universe (RESCEU), Graduate School of Science, The University of Tokyo, 7-3-1 Hongo, Bunkyo-ku, Tokyo 113-0033, Japan}
\affiliation{Department of Astronomy, Graduate School of Science, The University of Tokyo, Tokyo, Japan}

\author{Kotaro Fujisawa}
\affiliation{Department of Physics, Graduate School of Science, The University of Tokyo, 7-3-1 Hongo, Bunkyo-ku, Tokyo 113-0033, Japan}

\author{Toshikazu Shigeyama}
\affiliation{Research Center for the Early Universe (RESCEU), Graduate School of Science, The University of Tokyo, 7-3-1 Hongo, Bunkyo-ku, Tokyo 113-0033, Japan}
\affiliation{Department of Astronomy, Graduate School of Science, The University of Tokyo, Tokyo, Japan}

\begin{abstract}
We construct  axisymmetric self-similar solutions of transonic \textcolor{black}{outflow}s emanating from a point source including the effect of the rotation. The solutions are constructed exclusively on the equatorial plane. The features of solutions are determined by three parameters; the adiabatic index $\gamma$, the dimensionless coordinate of the transonic point, and the dimensionless azimuthal velocity at the transonic point. We classify the solutions into five groups according to the asymptotic behaviors. We find that the behaviors of the self-similar solutions change at $\gamma = 11/9$. In addition, some solutions show double-power-law density profiles, which are usually seen in ejecta from a binary merger or nova-like explosion. Thus, our self-similar solutions can be applied not only to the \textcolor{black}{outflow} blowing from the central spinning objects, but also to the ejecta erupted from the binary merger or nova-like explosion.
\end{abstract}


\section{Introduction} \label{sec:intro}
All objects in the universe rotate and have angular momentum. Among them, there are many objects formed via binary stellar mergers, for which the effects of rotation cannot be ignored. Some of such objects emit \textcolor{black}{outflow}s containing high angular momentum, and the \textcolor{black}{outflow}s are very important in the evolution of the system.

In addition to many objects that are spinning around and emitting rotational \textcolor{black}{outflow} at high velocities on their own, there are also objects that emit rotating ejecta by binary merger. Both observational researches and numerical simulations showed that binary mergers erupt a lot of such rotation-rich ejecta \citep[e.g.,][]{Tanikawa2015,Schneider2019,Shibata2019}, which bring essential information on the original objects. \cite{Schneider2019} reported three-dimensional simulations of the coalescence of two massive stars and showed there were some ejecta from the merger. Furthermore, binary white dwarf (WD) merger is known to cause a wide variety of phenomena such as type Ia supernovae \cite[e.g.,][]{Webbink1984,Iben1984,Bildsten2007,Fink2007,Guillochon2010,Dan2011,Pakmor2013}, the formation of a massive WD \cite[e.g.,][]{Schwab2016,Gvaramadze2019,Oskinova2020}, and collapse into a neutron star through accretion-induced collapse (AIC) \cite[e.g.,][]{Saio1985,Taam1986,Piro2013,Moriya2016}. The ejecta released from the binary WD merger is one of the important parameters that determines the physics of these phenomena. Some numerical results of binary WD mergers involving the ejecta have been reported \cite[e.g.,][]{Tanikawa2015,Sato2015,sato2016}. The ejecta from binary neutron star (NS) merger are very important because the ejecta become the electromagnetic counterpart known as kilonova \cite[e.g.,][]{Li1998,Metzger2010,Abbott2017} and a plausible origin of elements heavier than iron \citep{2014A&A...565L...5T,1974ApJ...192L.145L}. There are a lot of reports about the numerical simulations of binary NS merger and the properties of the ejecta \citep[for review, e.g.,][]{Shibata2019}. Some simulations showed the double-power-law density profile of the ejecta \citep[e.g.,][]{hotoke2018}.

Many stationary wind models have been constructed to understand the nature of these winds or ejecta blowing from the central objects without rotation \citep[e.g.,][]{1965SSRv....4..666P,kato1994} and with rotation \citep[e.g.,][]{Weber1967,2018arXiv181110777L,Kashiyama2019}.

\cite{1965SSRv....4..666P} gave a review of the one-dimensional steady wind model that includes the gravity of the central object. This model explains the solar wind in the vicinity of the sun very successfully. \cite{kato1994} constructed a steady wind model including the effects of radiation transfer and explicitly showed that the wind is accelerated by the radiation pressure. \citet{Weber1967} constructed an axisymmetric steady wind model including the central gravity, rotation, and magnetic field. The model is formulated as one-dimensional problems by focusing on the equatorial plane. \cite{Kashiyama2019}, by combining the models of \cite{Weber1967} and \cite{kato1994}, constructed an axisymmetric steady wind model including the central gravity, rotation, radiation transfer, and magnetic fields on the equatorial plane. This reported the nature of the wind from a fast rotating massive WD produced via binary merger, which is known to be blowing fast wind at a speed of up to $\sim16,000~\mathrm{km\,s^{-1}}$ from optical observations \citep[]{Gvaramadze2019}. The model attempts to reproduce the observed features of this object by the behavior of a rotating magnetic wind on the equatorial plane emitted from the central WD.

Recent short cadence observations will detect emissions from these events in the near future at a very early stage when the \textcolor{black}{outflow} has not yet reached a steady state. To prepare for this situation, 
we aim to construct a solution which can describe the early stage of the evolution of the \textcolor{black}{outflow} including the effect of rotation in this study.

In order to describe the evolution up to the steady state, a useful way is to assume the self-similarity of the evolving \textcolor{black}{blast waves} (so-called self-similar solution). \cite{1950RSPSA.201..159T}, \cite{1959sdmm.book.....S} and \cite{Parker1961} calculated self-similar \textcolor{black}{outflow} models to study a sudden ejection of matter such as a coronal mass ejection. Their models assumed the existence of a shock formed by a collision of the ejected material with the ambient medium. Thus, their self-similar models are strongly affected by the surroundings, and it is difficult to  understand the pure nature of the ejection. In addition, the self-similar solutions assuming the existence of one shock are effective only after the reverse shock disappeared. In order to describe the evolution of the \textcolor{black}{outflow} until a shock is generated, a transonic \textcolor{black}{outflow} model is an important tool. \cite{Cheng1977} and \cite{Fukue1984} reported such one-dimensional self-similar transonic \textcolor{black}{outflow} models. \cite{Cheng1977} constructed a self-similar transonic \textcolor{black}{outflow} model including the central gravity by assuming a polytropic equation of state. That is, the flow is assumed to be isentropic. 
\cite{Fukue1984} took an approach to the self-similar transonic \textcolor{black}{outflow} different from \cite{Cheng1977}. \cite{Fukue1984} did not adopt the polytropic equation of state but assumed the adiabatic evolution. None of these self-similar models took into account rotation. 

\cite{Kashiyama2019} reported that a rapid rotation plays an important role in the \textcolor{black}{outflow} properties. Thus, in order to apply to objects such as the fast-spinning WD, the transonic \textcolor{black}{outflow} model should be constructed by taking into account of rotation. Therefore, in this work we construct self-similar transonic \textcolor{black}{outflow} solutions including the central gravity and the high azimuthal velocity on the equatorial plane. It should be noted that the solutions constructed in this work include the solution of \cite{Cheng1977} because of the same setup except for rotation. Thus, we report the solutions with a topological examination including \cite{Cheng1977} solutions.




This paper is organized as follows. In Sect. \ref{sec:methods}, we introduce the self-similar setup and the integration of our work. In Sect. \ref{sec:result}, we present the results of our calculations and discuss them. In Sect. \ref{sec:conclusion}, we conclude the property of our self-similar calculations. 

\section{Methods} \label{sec:methods}
\subsection{Model}
We consider rotating stellar \textcolor{black}{outflow} emanating from a central point source that exerts the gravity. We assume that the system is axisymmetric around the rotational axis and consider the \textcolor{black}{outflow} exclusively on the equatorial plane. Under these assumptions, the continuity equation and the Euler equation become 
\begin{equation}
    \frac{\partial \rho}{\partial t}= -\frac{1}{r^2}\frac{\partial}{\partial r}(r^2\rho v_r),\label{eqn:basic1}
\end{equation}
\begin{equation}
     \frac{\partial v_r}{\partial t}= \frac{v_\phi^2}{r}-v_r\frac{\partial v_r}{\partial r}-\frac{1}{\rho}\frac{\partial p}{\partial r}-\frac{G\textcolor{black}{\textcolor{black}{M_{*}}}}{r^2},\label{eqn:basic2}
\end{equation}
\begin{equation}
    \frac{\partial v_\phi}{\partial t}=- \frac{1}{r}v_r\frac{\partial}{\partial r}(rv_\phi),\label{eqn:basic3}
\end{equation}
where, $r$ is the radius, $t$ is the time, $G$ is the gravitational constant, $\textcolor{black}{M_{*}}$ is the mass of the central object and $\rho,v_{r},v_{\phi},p$ are the density, the radial velocity, the azimuthal velocity, and the pressure, respectively. They are functions of $r$ and $t$. Here we assume that the flow is isentropic \textcolor{black}{because the cooling timescale is much longer than the dynamical timescale in the eruptive outflow context}:
\begin{equation}
    p = K\rho^\gamma,\label{eqn:adi}
\end{equation}
where $K$ is a constant, and $\gamma$ is the adiabatic index.

\subsection{Self-Similar Formalism}
We introduce the similarity variables as follows:
\begin{equation}
    x = \frac{r^3}{At^2},\label{eqn:dim1}
\end{equation}
\begin{equation}
    \rho = \left(\frac{r^2}{Kt^2}\right)^{\frac{1}{\gamma-1}}\Omega(x),
\end{equation}
\begin{equation}
    v_{r} = \frac{r}{t}V_{r}(x),
\end{equation}
\begin{equation}
    v_{\phi} = \frac{r}{t}V_{\phi}(x),\label{eqn:dim2}
\end{equation}
where $A = G\textcolor{black}{M_{*}}$.

Next, we rewrite equations (\ref{eqn:basic1})-(\ref{eqn:basic3}) in terms of similarity variables:
\begin{equation}
      W\vdot  df =   B,\label{eqn:Wdf}
\end{equation}
where $W$, $df$ and $B$ can be written as follows:
\begin{equation}
    W = \mqty(x \left(\gamma - 1\right) \Omega & 0 & x(\gamma-1)v \\
x(\gamma-1)v & 0 & x\gamma(\gamma-1)\Omega^{\gamma-2} \\
0 & x(\gamma-1)v  & 0),
\end{equation}
\begin{equation}
    df = \mqty(\frac{d V_r}{dx} \\
\frac{d V_\phi}{dx} \\
\frac{d \Omega}{dx} ),
\end{equation}
\begin{equation}
    B = \mqty(\frac{\left(- 3 \gamma V_{r}  + V_{r}  + 2\right) \Omega{ }}{3} \\
\frac{(1-\gamma)\left(V_{r}^{2}-V_{\phi}^{2}+ V_{r}+1/x \right)}{3}-\frac{2\gamma\Omega^{\gamma-1}}{3}\\
-\frac{(\gamma-1)(2V_r-1)V_\phi}{3}).
\end{equation}
Here, $v(x)$ has been introduced as
\begin{equation}
v(x)=V_{r}\left(x\right)-2/3.
\end{equation}
To solve these equations numerically, we rewrite the equation as follows:
\begin{equation}
  df =  W^{-1} \vdot  B.\label{eqn:df}
\end{equation}
To obtain a transonic solution, we require that $df$ should not diverge even at a critical point where the determinant of $W$ becomes $0$, i.e.,
\begin{equation}\label{eqn:detw}
  \mathrm{det}\ W = (\gamma-1)^3 x^3 \Omega^3 v\left(v^2-C^2\right) = 0,
\end{equation}
where $C=\sqrt{\gamma\Omega^{\gamma-1}}$ is the dimensionless sound speed. Otherwise, equation (\ref{eqn:df}) becomes singular at the critical point where $v=C$ and the flow truncates at this point. It should be noted from this equation that the density distribution can also be singular at a point where $v=0$, which describes a contact surface.

\subsection{Singularity Analysis and Boundary Conditions}\label{sec:boundary}

\begin{figure*}
 \centering
 \includegraphics[width=1.1\linewidth]{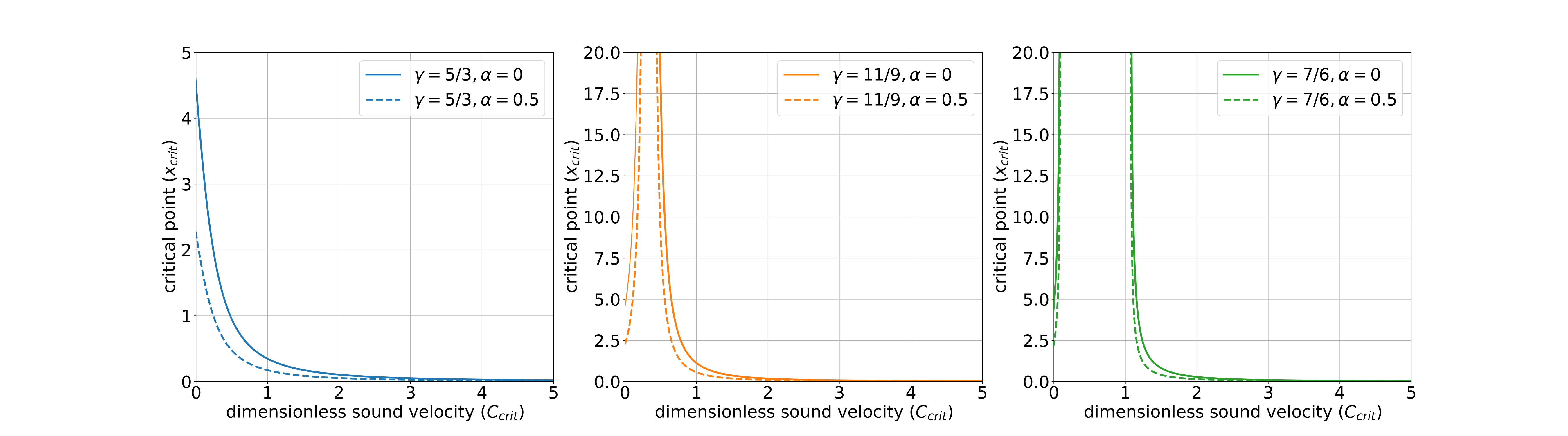}
\caption{The relation between the dimensionless sound velocity $C$ and $x_{\rm crit}$ at the transonic point for a few $\gamma$'s. From equation (\ref{eqn:critical2}), this relation has distinct features depending on whether the value of $\gamma$ is greater than $11/9$ or not. The three panel shows the three cases: $\gamma>11/9$ (left panel), $\gamma=11/9$ (middle panel) and $\gamma<11/9$ (right panel). For $\gamma >11/9$, $x_{\rm{crit}}$ has a maximum value, but for $\gamma \leq 11/9$, $x_{\rm{crit}}$ can take all positive values.}
 \label{fig:crit}
 \end{figure*}

In order to obtain a transonic solution from equation (\ref{eqn:Wdf}) without numerical divergence at the critical point, both of the denominator and numerator of equation (\ref{eqn:df}) must be equal to $0$, i.e. the following relation should be satisfied at the critical point:
\begin{equation}
  18C^2x+\frac{15\gamma-21}{\gamma-1}C x+2x-9+9V_\phi^2x=0.\label{eqn:critical}
\end{equation}
Here, we write $V_\phi\textcolor{black}{(x=x_{\rm crit})} = \alpha V_K\textcolor{black}{(x=x_{\rm crit})}$, where $V_K$ is the dimensionless azimuthal velocity in Kepler rotation and $\alpha$ is defined at the critical point. Since the azimuthal velocity in Kepler rotation is $v_K = \sqrt{G\textcolor{black}{M_{*}}/r}$,  $V_K$ can be expressed as follows:
\begin{equation}
  V_K = \frac{t}{r}\sqrt{\frac{G\textcolor{black}{M_{*}}}{r}} = \frac{1}{\sqrt{x}}.
\end{equation}
Substituting this to equation (\ref{eqn:critical}), we obtain the following relation at the critical point $x=x_{\rm{crit}}$:
\begin{equation}
  18C^2x+\frac{15\gamma-21}{\gamma-1}C x+2x-9(1-\alpha^2)=0\label{eqn:critical2},
\end{equation}
or
\begin{equation}
  x_{\rm crit}(\alpha,C) = \frac{9(1-\alpha^2)}{18C^2+\frac{15\gamma-21}{\gamma-1}C+2}.\label{eqn:crit}
\end{equation}
Figure \ref{fig:crit} shows this relation for several $\gamma$. It can be easily seen from the equation and the figure that the position of the critical point decreases when $|\alpha|$ (the rotation) is large for a constant $C$. Here, let $f(C)$ be the denominator of equation \eqref{eqn:crit}, the discriminant $D$ of the quadratic equation $f(C) = 0$ :
\begin{equation}
    D = \frac{9\left(9\gamma-11\right)\left(\gamma-3\right)}{(\gamma-1)^2},
\end{equation}
indicates that $f(C)$ and equation (\ref{eqn:crit}) change the behavior depending on $\gamma$ greater or less than $11/9$. When $\gamma>11/9$, $D$ becomes negative and $f(C)$ does not vanish for any real $C$. Thus, $x_{\rm{crit}}$ takes the maximum value $x_{\rm{crit}}=9(1-\alpha^2)/2$ at $C=0$, decreases monotonically with increasing $C$, and approaches to $0$ as $C$ approaches infinity (left panel of Fig. \ref{fig:crit}).  
When $\gamma \leq 11/9$, $x_{\rm{crit}}$ can take all positive values because $f(C) = 0$ has a real solution of $C$ and then $x_{\rm{crit}}$ can be infinite (center and right panels in Fig. \ref{fig:crit}).

From equations (\ref{eqn:detw}) and (\ref{eqn:critical2}), all the physical quantities at the singular point are obtained by giving $x_{\rm{crit}}$ and $\alpha$. Therefore, we set the values of $x_{\rm{crit}}$ and $\alpha$ and initial values at the critical point and integrate equations (\ref{eqn:Wdf}) inwardly and outwardly avoiding numerical divergence. 
\subsection{Integration}
We use the Runge–Kutta method to numerically integrate equation (\ref{eqn:df}) from the critical point, using the \verb|scipy.integrate| module (\citealt{2020SciPy-NMeth}). As mentioned in Sect.\ref{sec:boundary}, the integration starts from points near the critical point for some parameter sets of $(\gamma,x_{\mathrm{crit}},\alpha)$.

\section{Results and Discussion} 

In this section, we present self-similar solutions for several parameter sets and the analysis of the self-similar solutions. In addition, we present the solutions with dimensions and discuss the features of the solutions.
\subsection{Dimensionless solutions}\label{sec:result}
\begin{figure*}
 \centering
 \includegraphics[width=1.1\linewidth]{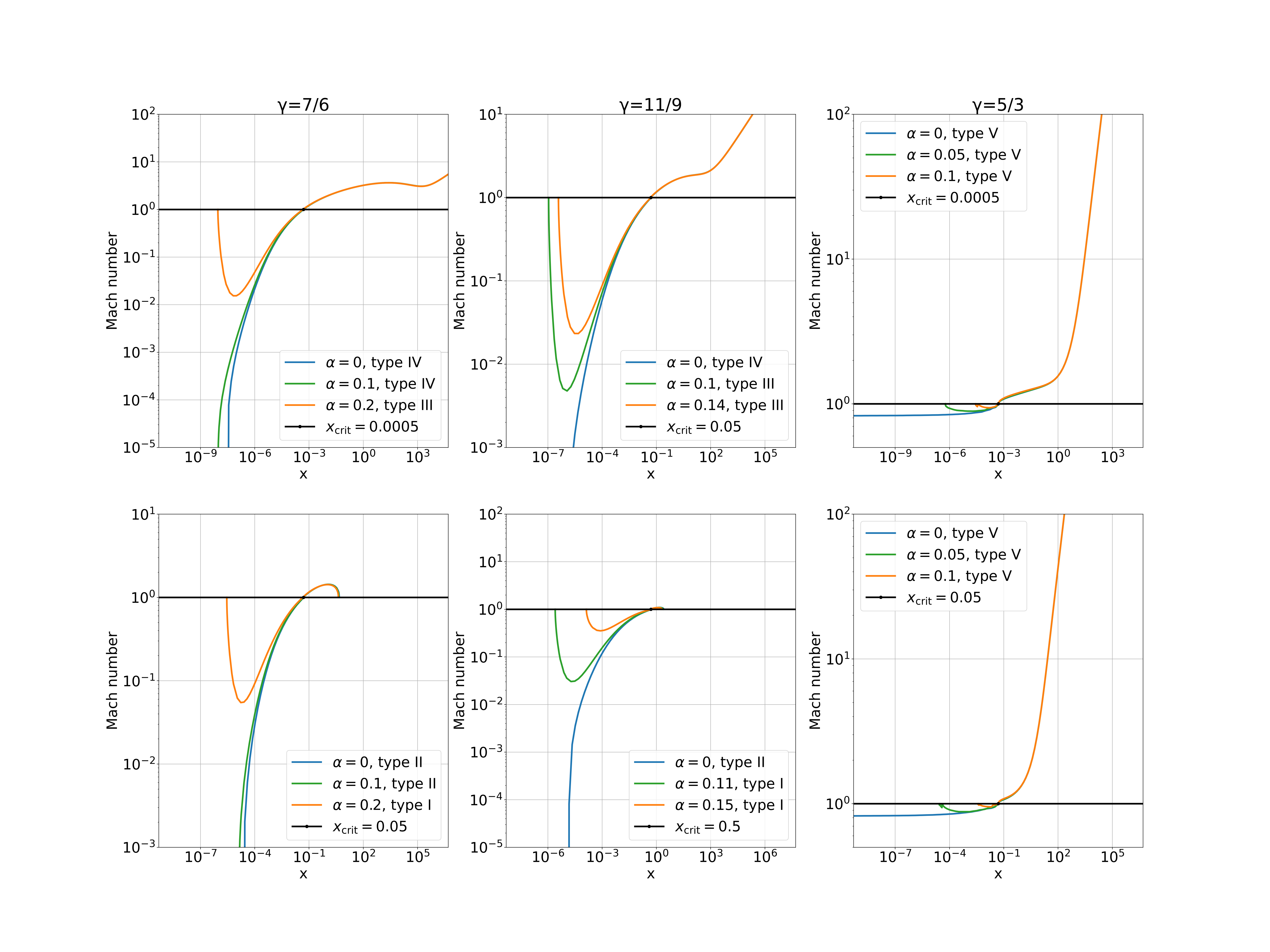}
\caption{The dimensionless Mach numbers as functions of $x$ for several parameter sets. Some solutions truncate at the point where $v=C$ (i.e., $M=1$) again inside and/or outside of the transonic point $x=x_{\rm{crit}}$. The other solutions truncate at the point where $v=0$ or $M=0$ inside of the transonic point. The left two panels show the distribution of $M(x)$ for $\gamma = 7/6$. The middle two show that for $\gamma = 11/9$ and the right two show that for $\gamma = 5/3$. The upper and lower panels show solutions with different coordinates $x_{\rm{crit}}$ of transonic points.}
 \label{fig:non_dim_mach}
 \end{figure*}

Figure \ref{fig:non_dim_mach} shows the distribution of the dimensionless Mach number defined as 
\begin{equation}
    M(x) = \frac{v(x)}{C(x)} = \frac{V_r(x)-2/3}{\sqrt{\gamma\Omega^{\gamma-1}(x)}},
\end{equation}
for each parameter set $(\gamma,x_{\mathrm{crit}},\alpha)$.
Where $M(x)=1$ (the sonic point) at a place other than $x=x_{\rm crit}$, the derivatives in equation (\ref{eqn:df}) diverge and the flow truncates as shown in Figure \ref{fig:non_dim_mach}. This means that the flow is in contact with the vacuum at the sonic singular point \textcolor{black}{with some exceptions (see section \ref{sec:rotation})}.

\subsubsection{Classification}
We found that solutions with rotation ($\alpha\neq0$) can be classified into five types based on the number of sonic points:  solutions with $\gamma\leq 11/9$ being classified into four types (\textcolor{black}{I, II, III, and IV}) and solutions with $\gamma>11/9$ into one (\textcolor{black}{V}). 
 The flow with $\gamma \leq 11/9$ has a sonic point or a contact surface inside the transonic point depending on whether $|\alpha|$ is larger than a threshold value or not. For $\gamma > 11/9$ (type \textcolor{black}{V}), on the other hand, an inner sonic point always appears unless the flow does not rotate at all (i.e., $\alpha\neq0$).  

\subsubsection{Solutions with $\gamma \leq 11/9$}
\begin{figure*}
 \centering
 \includegraphics[width=1.1\linewidth]{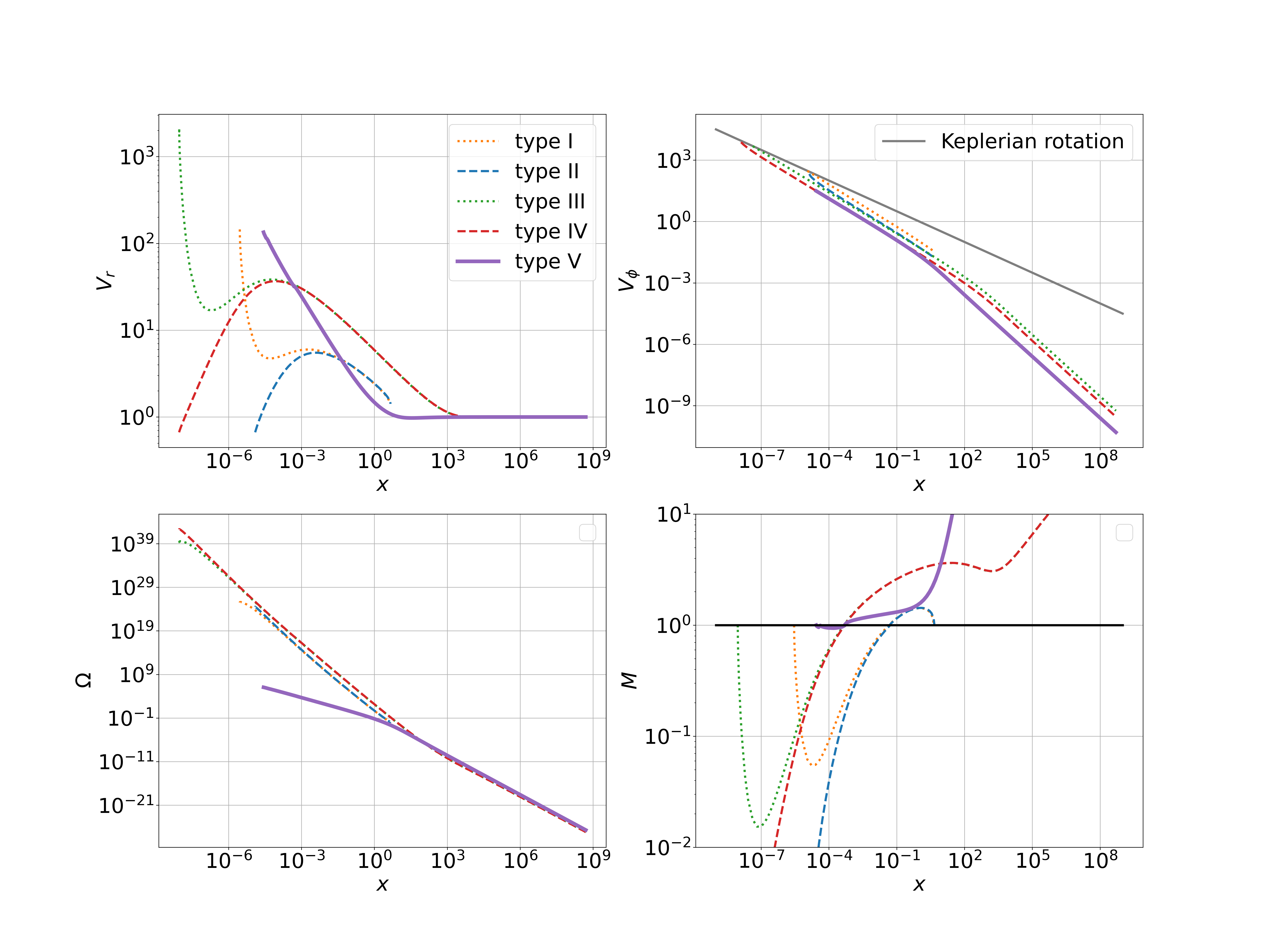}
\caption{A comparison of dimensionless variables as functions of $x$ between different types of solutions. In this figure, the following parameter sets $(\gamma,x_{\mathrm{crit}},\alpha)$ are used: type \textcolor{black}{I} (7/6,0.05,0.2), type \textcolor{black}{II} (7/6,0.05,0.1), type \textcolor{black}{III} (7/6,0.0005,0.2), type \textcolor{black}{IV} (7/6,0.0005,0.1) and type \textcolor{black}{V} (5/3,0.0005,0.1).}
 \label{fig:A}
 \end{figure*}

For $\gamma \leq 11/9$, the solutions are classified based on whether a singular point exists inside ($x<x_{\rm crit}$) or outside ($x>x_{\rm crit}$) of the transonic point. Type \textcolor{black}{I} solution has sonic points on both sides of the transonic point  (see the line with a parameter set of $(\gamma,x_{\mathrm{crit}},\alpha)=(7/6,0.05,0.2)$ in Fig.\,\ref{fig:non_dim_mach}, for example) and the flow truncates on both sides of the sonic points. Type \textcolor{black}{II} solution truncates at the sonic point outside the transonic point and also truncates at the contact surface inside of the transonic point where $V_r=2/3$ ($v=0$) instead of $M=1$ (e.g., $(\gamma,x_{\mathrm{crit}},\alpha)=(7/6,0.05,0.1)$).  On the other hand, type \textcolor{black}{III} solution has an inner sonic point but no outer sonic point and extends to infinity (e.g., $(\gamma,x_{\mathrm{crit}},\alpha)=(7/6,0.0005,0.2)$). Type \textcolor{black}{IV} solution (e.g., $(\gamma,x_{\mathrm{crit}},\alpha)=(7/6,0.0005,0.1)$) has a contact surface inside the transonic point and has no sonic point on either side of the transonic point. At the inner sonic point in type \textcolor{black}{I} and \textcolor{black}{III} solutions, $V_r$ and $\Omega$ diverge to infinity as shown in Figure \ref{fig:A}. Type \textcolor{black}{II} and type \textcolor{black}{IV} solutions appear when the value of $|\alpha|$ is small or the rotation at the transonic point is slow. 

Solutions of types \textcolor{black}{III} and \textcolor{black}{IV} extend to infinity and approach homologous expansion. This is indicated from Figure \ref{fig:A} in which $V_r$ of these solutions approach unity at large $x$. We recognize such flows approaching homologous expansion ($V_r = 1$) as \textcolor{black}{outflow} type solutions.  We can obtain the asymptotic behaviors of the density and the azimuthal velocity listed in table \ref{tab} by substituting $V_r = 1$ ($v = 1/3$) into the equation (\ref{eqn:Wdf}). Equations of the conservation of the mass and the angular momentum yield
\begin{equation}
     \frac{x}{3}\frac{d\Omega}{dx} = -\Omega,
\end{equation}
and
\begin{equation}
   \frac{x}{3} \frac{dV_\phi}{dx} = - \frac{V_\phi}{3},
\end{equation}
respectively. Then $\Omega\propto x^{-3}$ and $V_\phi\propto x^{-1}$ satisfy these equations. 
In the inner region where $x_{\rm{min}}\ll x\ll x_{\rm{crit}}$, $V_r\gg 1$ from Figure \ref{fig:A}. Then we can show $V_\phi\propto x^{-2/3}$ because 
\begin{equation}
    xV_r  \frac{dV_\phi}{dx} = \frac{2V_\phi}{3},
\end{equation}
 from equation (\ref{eqn:df}).

All solutions with $\gamma\leq 11/9$ and a finite $\alpha$ have rotational velocities greater than the Kepler rotation in the innermost region. Note that the boundary surfaces between the four types (\textcolor{black}{I, I, III and IV}) in the parameter space of $(\gamma,x_{\mathrm{crit}},\alpha)$ can only be determined  numerically. 

\subsubsection{Solutions with $\gamma>11/9$}
Type \textcolor{black}{V} solution with $\gamma>11/9$ always has an inner sonic point but no outer sonic point. As shown in Table \ref{tab}, the asymptotic power-law behaviors of physical quantities have the same exponents as types \textcolor{black}{III} and \textcolor{black}{IV}, but there is a difference from type \textcolor{black}{III} solutions. Type \textcolor{black}{V} without rotation extends to the center, while type \textcolor{black}{III} never reaches the center. 

The fact that the behavior of the solution changes across $\gamma=11/9$ is consistent with the change in behavior across $n=1/(\gamma-1)=9/2$ in equation (A11) in \cite{Cheng1977}. 

\subsubsection{Outer critical points}
\begin{figure*}
 \centering
 \includegraphics[width=\linewidth]{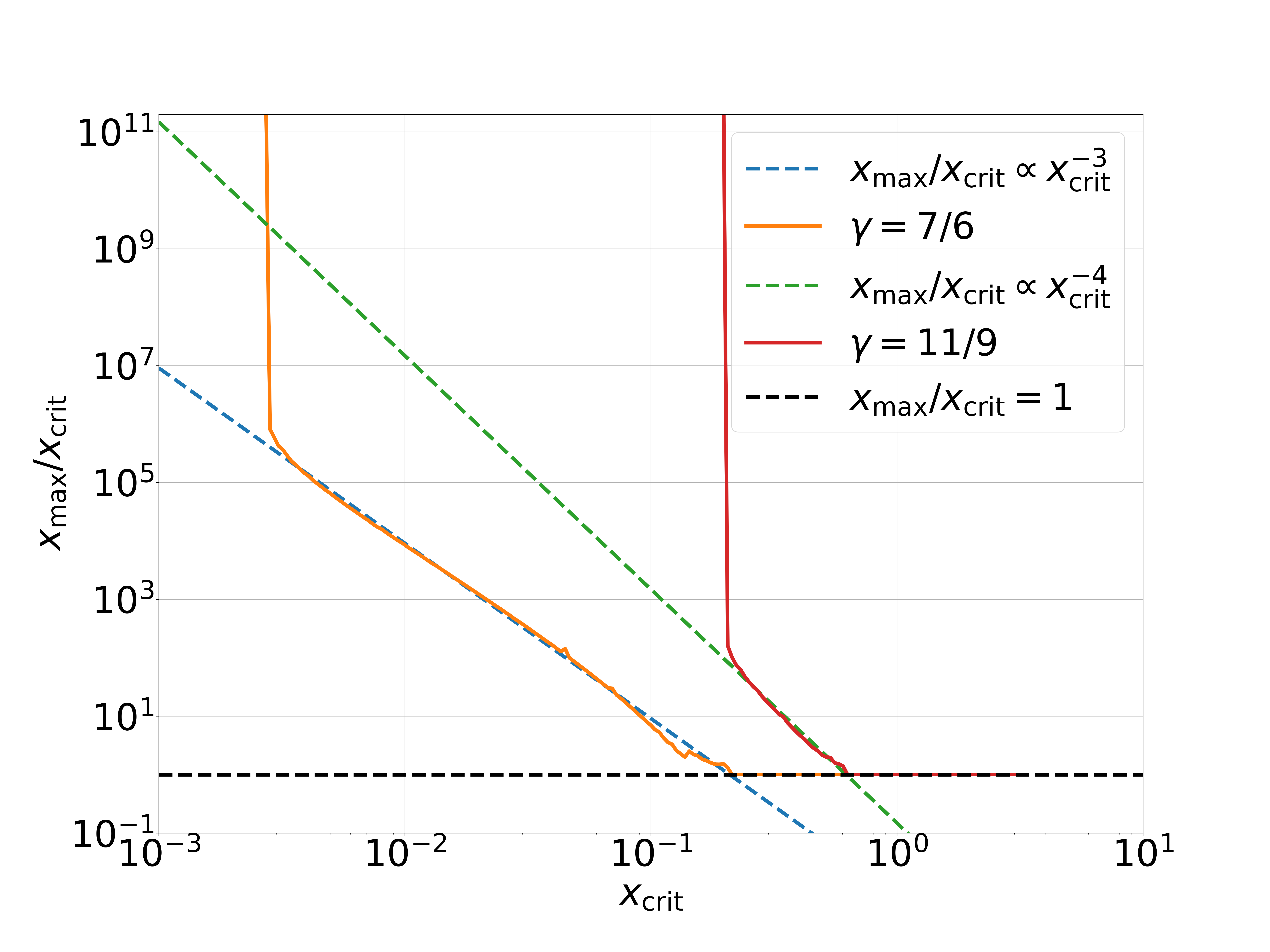}
\caption{The relation between $x_{\rm{crit}}$ and $x_{\rm{max}}/x_{\rm{crit}}$ for two different adiabatic indices denoted in the legend. Here, $x_{\rm{max}}$ denotes the coordinate of the outer sonic point. Both of the lines show the type of the flow changes from \textcolor{black}{II} to \textcolor{black}{IV} with decreasing $x_{\rm{crit}}$. A large $x_{\rm{crit}}$ inhibits the integration toward the outer region and thus results in $x_{\rm{max}}$ very close to $x_{\rm{crit}}$. Note that the dashed horizontal line indicates $x_{\rm max}/x_{\rm crit}=1$.}
 \label{fig:x_effect}
 \end{figure*}

Figure \ref{fig:x_effect} exemplifies transitions from type \textcolor{black}{II} to type \textcolor{black}{IV} solutions with decreasing $x_{\rm crit}$ by showing the relation between the critical point $x_{\rm{crit}}$ and the ratio of the coordinates of the outer sonic point $x_{\rm{max}}$ to $x_{\rm{crit}}$. The divergence of the values of $x_{\rm{max}}/x_{\rm{crit}}$ to infinity in this plot indicates that the type of the solution shifts from type \textcolor{black}{II} to type \textcolor{black}{IV}. 
Another noticeable feature observed from Figures \ref{fig:non_dim_mach} and \ref{fig:x_effect} is that the ratio $x_{\rm{max}}/x_{\rm crit}$ decreases with increasing $x_{\rm{crit}}$ following a scaling law:  $x_{\rm{max}}/x_{\rm{crit}}\propto x_{\rm{crit}}^{18(1-\gamma)}$ as indicated in Figure \ref{fig:x_effect}. 

\subsubsection{Inner critical points}
\begin{figure*}
 \centering
 \includegraphics[width=\linewidth]{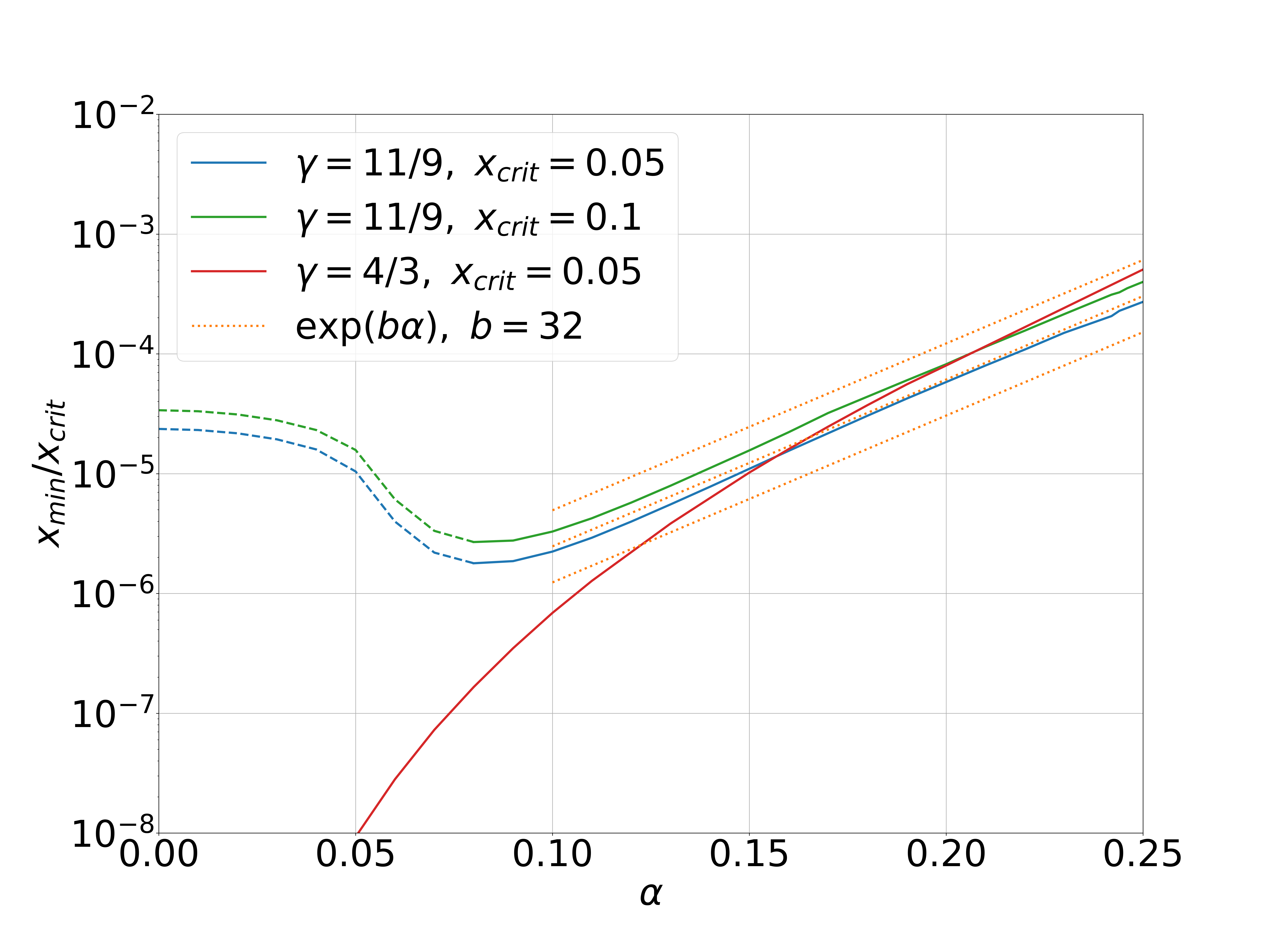}
\caption{The relation between the rotation strength $\alpha$ and $x_{\rm{min}}/x_{\rm{crit}}$. Here, $x_{\rm{min}}$ denotes the coordinate of the inner truncation point. The flow truncates because $v = C$ (solid lines) or $v = 0$ (dashed lines) at $x=x_{\rm{min}}$. For the truncation with $v=C$, $x_{\rm{min}}$ is found to monotonically increase with increasing rotation ($|\alpha|$). For large $\alpha$, $x_{\rm{min}}/x_{\rm{crit}}\propto\exp(b\alpha)$ is satisfied as far as $\gamma \leq 11/9$. When the parameter set is $(\gamma,x_{\rm{crit}})=(7/6,0.05)$, the value of $b$ becomes about 28.}
 \label{fig:rot_effect}
 \end{figure*}

Figures \ref{fig:non_dim_mach} and \ref{fig:rot_effect} show that the faster the rotation (the larger $|\alpha|$), the closer the distance between the transonic point $x_{\rm{crit}}$  and the inner sonic point for the flow with $\gamma > 11/9$. Thus the ratio $x_{\rm{min}}/x_{\rm{crit}}$ increases with increasing $|\alpha|$ and type \textcolor{black}{V} solution with a finite value of $\alpha$ has an inner sonic point (at $x=x_{\rm{min}}$) and the flow truncates there. On the other hand, the flow with $\gamma \leq 11/9$ has a different trend because the flow with a small $|\alpha|$ can truncate at the contact surface where $v$ becomes 0 rather than at a sonic point. When the inner boundary is a contact surface (i.e., $v=0$), the ratio of $x_{\rm{min}}/x_{\rm{crit}}$ decreases with increasing $\alpha$ (dashed line in Fig. \ref{fig:rot_effect}). In contrast, the flow with a larger $|\alpha|$ truncates at the point where $v=C$, and the ratio $x_{\rm{min}}/x_{\rm{crit}}$ increases with increasing $\alpha$. This trend can be approximated as $x_{\rm{min}}/x_{\rm{crit}}\propto \exp(b|\alpha|)$, where $b$ is a proportional coefficient. 
 For solutions with an inner sonic point, the values of $x_{\rm{min}}/x_{\rm{crit}}$ monotonically decrease with increasing $x_{\rm{crit}}(\alpha)$. Since $x_{\rm{crit}}(\alpha)$ is a monotonically decreasing function of $\alpha$, the subsonic region shrinks with increasing $|\alpha|$. This feature can be seen in the steady \textcolor{black}{outflow} model as well \citep[e.g., see figure 3 in][]{2018arXiv181110777L}. In addition, the value of $x_{\rm{min}}/x_{\rm{crit}}$ increases with decreasing $x_{\rm{crit}}$, which means that the flow truncates immediately close to the transonic point for small $x_{\rm{crit}}$. 
These two facts reflect the relative importance of the gravity and the rotation. A small $x_{\rm{crit}}$ indicates a significant centrifugal force relative to the gravity since $x={r^3}\big/{G\textcolor{black}{M_{*}}t^2}$. 
\subsubsection{The effect of rotation}\label{sec:rotation}
 Figure \ref{fig:A} compares distributions of the dependent variables for different types of solutions. In each type of solutions, the flow truncates at a point inside the transonic point with $M=0$ or $M=1$, and  the azimuthal velocity there sometimes exceeds that of the Kepler rotation where the flows truncate. Furthermore, since $V_\phi$ becomes much smaller than $V_r$ and the Kepler rotation outside the transonic point ($x>x_{\rm{crit}}$), there is little effect of rotation in any type of solution.

 \subsection{Self-Similar Solutions with Dimension}
In this section, we attach the solutions obtained in Sect. \ref{sec:result} with dimensions by using equations \eqref{eqn:dim1} - \eqref{eqn:dim2}. For simplicity, we show solutions with dimensions setting $A=1$ and $K=1$. 
We have classified solutions into five types. Both of type \textcolor{black}{I} and type \textcolor{black}{II} solutions terminate at a certain point outside the critical point. Type \textcolor{black}{III} and type \textcolor{black}{IV} solutions have density distributions increasing with radius. This feature would not be feasible in reality. Furthermore, some of these \textcolor{black}{I-V} types of solutions rotate faster than the Kepler rotation inside the transonic point. Such a high azimuthal velocity may not be feasible without taking into account the effects of strong magnetic fields \citep[e.g.,][]{Kashiyama2019}. 

Thus, we focus on type \textcolor{black}{V} solutions in which the azimuthal velocity never exceeds that of the Kepler rotation and show their features in the rest of this section. In the limit of large $r$ or small $t$, the velocities and the density of type \textcolor{black}{V} solution have the following asymptotic temporal and radial dependence;
\begin{equation}
  v_r(r,t)\propto \frac{r}{t},\label{eqn:vr}
\end{equation}
\begin{equation}
  v_\phi(r,t)\propto \frac{t}{r^2},
\end{equation}
\begin{equation}
  \rho(r,t)\propto t^{\frac{6\gamma-8}{\gamma-1}}r^{-\frac{9\gamma-11}{\gamma-1}}\propto t^{-3}\left(\frac{r}{t}\right)^{-\frac{9\gamma-11}{\gamma-1}}.\label{eqn:rho}
\end{equation}
Equations (\ref{eqn:vr}) and (\ref{eqn:rho}) show that matter in the outer region freely expands.  

In the inner region ($x_{\rm min}\ll x\ll x_{\rm crit}$), the velocities and the density have the following asymptotic temporal and radial dependences;
\begin{equation}
  v_r(r,t)\propto t^{\frac{3\gamma-5}{3\gamma-3}}r^{-\frac{2\gamma-3}{\gamma-1}},\label{eqn:v_r}
\end{equation}
\begin{equation}
  v_\phi(r,t)\propto t^{1/3}r^{-1},\label{eqn:v_phi}
\end{equation}
\begin{equation}
  \rho(r,t)\propto t^0r^{-\frac{1}{\gamma-1}}.\label{eqn:equ}
\end{equation}
Equations (\ref{eqn:v_r}) and (\ref{eqn:equ}) indicate that the solution in the inner region is in a nearly hydrostatic equilibrium state as long as $\gamma<5/3$ and $v_\phi\ll v_K$, because the radial velocity becomes much smaller than the sound speed there. Thus,
\begin{equation}
  \gamma K\rho^{\gamma-2}\frac{d\rho}{dr}+\frac{G\textcolor{black}{M_{*}}}{r^2}=0,
\end{equation}
holds there. However, in Type \textcolor{black}{V} solution, it is numerically found that if the $|\alpha|$ is smaller than a certain threshold, the azimuthal velocity exceeds the Kepler velocity in the inner region, and this $|\alpha|$ threshold becomes smaller for larger $\gamma$. For example, $\gamma = 1.5$ yields the threshold of $|\alpha| \sim 0.1$ and $|\alpha| \sim 0.007$ for $\gamma = 1.6$. Thus, type \textcolor{black}{V} solutions with lower $|\alpha|$ rotate faster than the Kepler rotation at the innermost points while with higher $|\alpha|$ do not exhibit such peculiar features. Thus, type \textcolor{black}{V} solution has the inner region that can be regarded as a rotating steady state as long as $v_\phi\ll v_K$ holds. The steady state can be attained with large $t$'s as well as with small $x$'s because a large $t$ also indicates a small $x$. Thus the inner region of type \textcolor{black}{V} solution attains a rotating steady state and this region expands over time to a larger radius. \textcolor{black}{In this type V solutions, the total angular momentum increases over time and the angular momentum continuously enters from the inner boundary.}

The above calculations show that the flow with $\gamma>11/9$ converges to homologous expansion at large $x$, and its density profile has a double-power-law distribution as follows:
\begin{equation}
 \rho(r,t)\propto r^{-\frac{1}{\gamma-1}}\ \rm{(inner)},
\end{equation}
\begin{equation}
 \rho(r,t)\propto t^{\frac{6\gamma-8}{\gamma-1}}r^{-\frac{9\gamma-11}{\gamma-1}}\ \rm{(outer)}.
\end{equation}

Since the double-power-law of the density profile is realized when $\gamma>11/9$, the density profile is always double-power-law when the pressure is dominated by radiation ($\gamma\sim4/3$).
Since such double-power-law of the density profile is commonly observed in ejecta erupted from binary mergers \citep[e.g.,][]{Shibata2019}, the model constructed in this work may explain the ejecta from the binary-merger system \textcolor{black}{when the ejecta mass can be ignored compared with the central object mass}. Note that, the density profile becomes a "single-power law" when $\gamma = 4/3$. 
\textcolor{black}{In this work, we assume that the ejecta mass can be neglected, and this assumption holds when}
\begin{equation} \label{eqn:mass_condition}
\textcolor{black}{
    M_* \gg 4\pi\int r^2\rho dr
    }
\end{equation}
\textcolor{black}{
holds if we assume the spherical symmetry. Using our solutions, we can calculate the right hand side:
}
\begin{equation}
\textcolor{black}{
    4\pi\int r^2\rho dr = 4\pi\int\left(At^2x\right)^{2/3}\left(\frac{r^2}{Kt^2}\right)^{\frac{1}{\gamma-1}}\Omega(x)dr.
    }
\end{equation}
\textcolor{black}{
With fixed time $t$, this equation can be expressed as 
}
\begin{equation}
\textcolor{black}{
    4\pi\int r^2\rho dr =A^{\frac{3\gamma-1}{3\gamma-3}}K^{-\frac{1}{\gamma-1}}t^{\frac{6\gamma-8}{3\gamma-3}}\int x^{\frac{2}{3\gamma-3}}\Omega(x)dx.
    }
\end{equation}
\textcolor{black}{
Thus, for the solutions with $\gamma > 4/3$, this assumption holds at small $t$, and the solutions with $\gamma < 4/3$, this holds at large $t$. However, for the solutions with $\gamma < 4/3$, the ejecta mass is monotonically decreasing and it is not an outflow solution. In type V solution with $(\gamma,x_{\mathrm{crit}},\alpha)= (5/3,0.05,0.1)$, for example, this relation can be calculated as 
}
\begin{equation}\label{eqn:ejectamass}
\textcolor{black}{
    4\pi\int r^2\rho dr \sim 1.4 A^{2}K^{-3/2}t.
    }
\end{equation}
\textcolor{black}{
In the context of massive white dwarf observed by \cite{Gvaramadze2019}, $M_*\sim1.5 M_\odot$ and the mass loss rate $\dot{M}\sim3.5\times10^{-6}M_\odot$/yr are estimated by \cite{Gvaramadze2019} and \cite{Kashiyama2019}. If the adiabatic index is $\gamma = 5/3$, the adiabatic constant can be estimated as $K\sim 4\times10^{31}$ by differentiating equation (\ref{eqn:ejectamass}) with $t$. Substituting these values to equation (\ref{eqn:mass_condition}), we can obtain a criterion that the self-gravity of ejecta mass can be neglected as
}
\begin{equation}
\textcolor{black}{
    t\ll4\times10^5 \mathrm{yr}.
    }
\end{equation}
\textcolor{black}{
This is much larger than the evolution timescale of the massive white dwarf system which is estimated to be a few $10^3$ year \citep[][]{Kashiyama2019}. Thus this work may explain the outflow evolution.
}

\begin{figure*}
 \centering
 \includegraphics[width=1.1\linewidth]{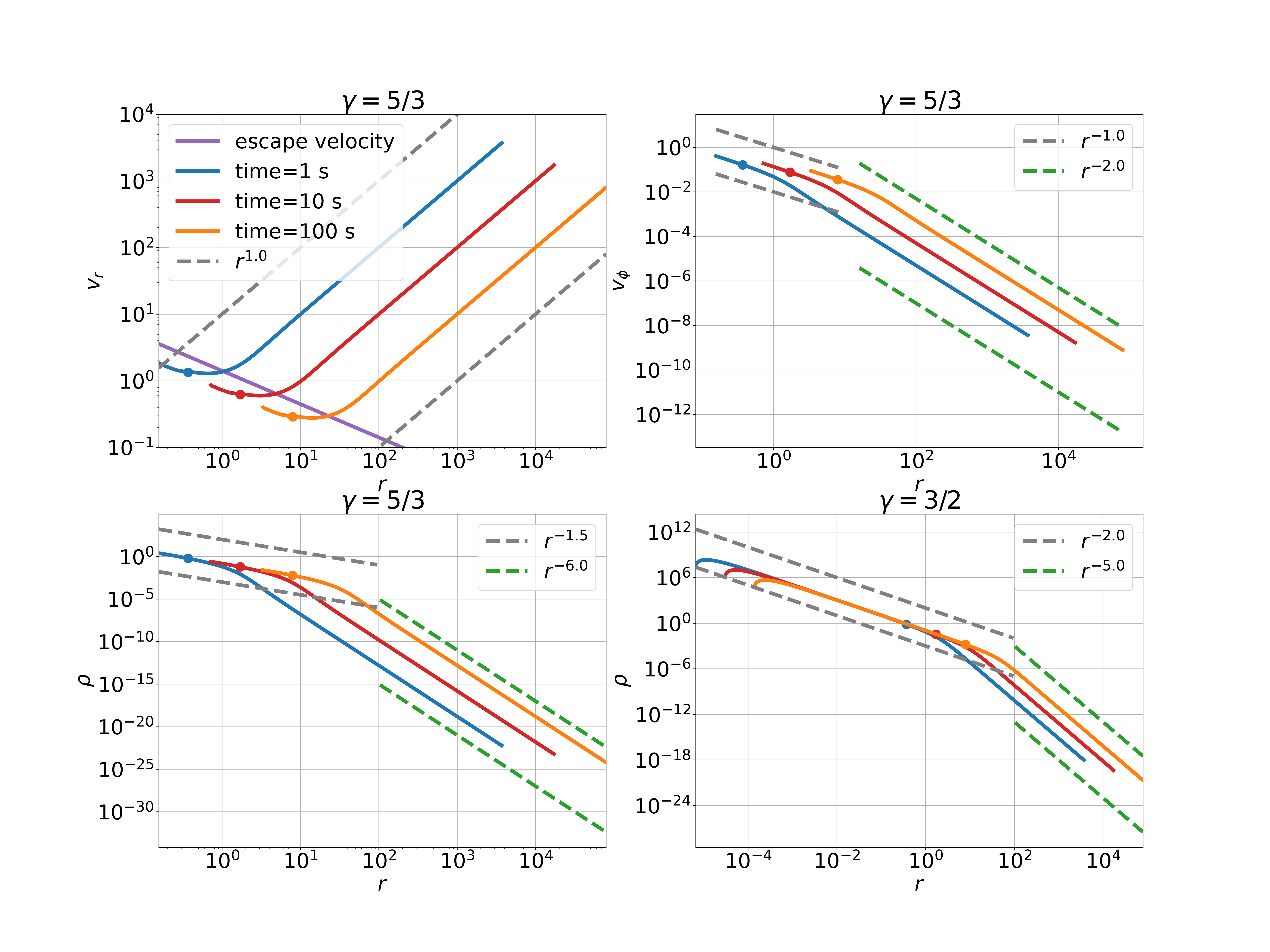}
\caption{Time evolution of physical quantities of type \textcolor{black}{V} solutions. The upper two panels and the bottom left panel show the time evolution of  the radial velocity (the top left), the azimuthal velocity (the top right), and the density (the bottom left) of type \textcolor{black}{V} solution   $(\gamma,x_{\mathrm{crit}},\alpha)= (5/3,0.05,0.1)$ where the azimuthal velocity does not exceed the Kepler rotation anywhere. The lower right panel shows the time evolution of the density of type \textcolor{black}{V} solution  $(\gamma,x_{\mathrm{crit}},\alpha)= (3/2,0.05,0.005)$ where the azimuthal velocity exceeds the Kepler rotation (the bottom right panel) in the innermost region. Each dot indicates a transonic point where $x=x_{\rm{crit}}$. For the former solution, quasi-hydrostatic equilibrium is established because the flow velocity is subsonic in the inner region. For the latter solution, the rotational effect is not negligible and the radial velocity approaches the sound speed in the inner region.}
 \label{fig:dimE}
 \end{figure*}

Moreover, Figure \ref{fig:dimE} shows that the radial velocity is below the escape velocity in the inner region. 
By contrast, matter in the outer most region is freely expanding. The double power law in the density distribution is a result of a "weak" explosion in which the explosion energy is a fraction of the binding energy of the original matter. 
 Ejecta with such a feature have been observed from numerical simulations for the eruption of the envelope of a massive star prior to a supernova \cite{Tsuna2021}. 
 Thus there is potential for applications in the formation of the circumstellar matter of an interaction-powered supernova even if there is rotation.
 
 \textcolor{black}{In this work, we only consider the outflows on the equatorial plane, however, we can use the same basic equations \eqref{eqn:basic1}-\eqref{eqn:basic3} to study the angular dependency of the outflows (e.g., \citealp{Muller2014}). \citet{Muller2014} provided two-dimensional outflow models with rotation by assuming zero angular flow velocity. Suppose there is no interaction between radial flows with different polar angles. In that case, it is possible to investigate the properties of rotating outflows in a two-dimensional plane with each polar angle. Those are future works.}






%

\section{Conclusions} \label{sec:conclusion}
In this work, we have constructed a rotating transonic self-similar \textcolor{black}{outflow} in the equatorial plane. We find that \textcolor{black}{outflow} solutions exist in the case of $\gamma > 11/9$, and these solutions have double-power-law density profiles as follows:
\begin{equation}
 \rho(r,t)\propto r^{-\frac{1}{\gamma-1}}\rm{(inner)},
\end{equation}
\begin{equation}
 \rho(r,t)\propto t^{\frac{6\gamma-8}{\gamma-1}}r^{-\frac{9\gamma-11}{\gamma-1}} \rm{(outer)}.
\end{equation}
Such double-power-law density profiles are also common in the ejecta erupted from binary merger systems. 

The effect of rotation is noticeable in the subsonic region. In the subsonic region, the inner singular point appears and the location of the point approaches the transonic point as the rotation increases. This behavior is consistent with the rotating steady \textcolor{black}{outflow}. On the other hand, the rotation does not affect the solution of the supersonic region. Though our solutions have some common features to steady rotating \textcolor{black}{outflow} models, our solutions never reach any steady states and continue to expand because of the existence of the inner sonic point.

The self-similar solutions in this paper do not include the effect of the magnetic fields. Magnetic fields are keys to understanding an \textcolor{black}{outflow} from merger remnant \citep[][]{Kashiyama2019,Schneider2019}. In order to study an \textcolor{black}{outflow} from a fast rotating white dwarf with strong magnetic fields, it is necessary to construct a self-similar \textcolor{black}{outflow} model with magnetic fields, which will be our future works.

\begin{acknowledgments}
This work is also supported by JSPS KAKENHI grant Nos. 20K14512 (KF), 20H05639, 22K03688, and 22K03671 (TS), MEXT, Japan.
\end{acknowledgments}
\bibliography{reference}{}
\bibliographystyle{aasjournal}

 \begin{deluxetable*}{c|ccc}\label{tab}
\tablenum{1}
\tablecaption{The asymptotic forms of the dimensionless variables $\Omega$ and $V_\phi$ for each type. In types \textcolor{black}{I} and \textcolor{black}{II}, the exponent is for $\gamma=7/6$. In types \textcolor{black}{III}, \textcolor{black}{IV} and \textcolor{black}{V}, the exponent is for general $\gamma$. While type \textcolor{black}{III}, \textcolor{black}{IV}, and \textcolor{black}{V} solutions exhibit the same asymptotic behavior of the dimensionless variables, the density distributions ($\rho(r,t)$) of these types are different because of different $\gamma$. It should be noted that when the flow does not truncate at $x > x_{\rm{crit}}$, we set $x_{\rm{max}}=\infty$.}
\tablewidth{0pt}
\tablehead{
\colhead{Type} &\colhead{Asymptotic forms for $x_{\rm{min}}\ll x\ll x_{\rm{crit}}$} & \colhead{Asymptotic forms for $ x_{\rm{crit}} \ll x\ll x_{\rm{max}}$} &\colhead{Examples of the parameter sets}
}
\startdata
\textcolor{black}{I} & $V_\phi \propto x^{-0.68}$ & $V_\phi \propto x^{-0.72}$&$(\gamma,x_{\mathrm{crit}},\alpha)$\\
&$\Omega\propto x^{-4.9}$&$\Omega\propto x^{-4.2}$ &$=(7/6,0.05,0.2)$\\
&&& \\
\textcolor{black}{II} & $V_\phi \propto x^{-0.72}$ & $V_\phi \propto x^{-0.72}$&$(\gamma,x_{\mathrm{crit}},\alpha)$\\
&$\Omega\propto x^{-4.9}$&$\Omega\propto x^{-4.2}$&$=(7/6,0.05,0.1)$\\
&&& \\
\textcolor{black}{III, IV} & $V_\phi \propto x^{-2/3}$ & $V_\phi \propto x^{-1}$&$(\gamma,x_{\mathrm{crit}},\alpha)$\\
&$\Omega\propto x^{-\frac{1}{\gamma-1}}$&$\Omega\propto x^{-3}$&$=(7/6,0.0005,0.2)$\\
&&& \\
\textcolor{black}{V} & $V_\phi \propto x^{-2/3}$ & $V_\phi \propto x^{-1}$&$(\gamma,x_{\mathrm{crit}},\alpha)$\\
&$\Omega\propto x^{-\frac{1}{\gamma-1}}$&$\Omega\propto x^{-3}$&$=(5/3,0.0005,0.1)$\\
\enddata

\end{deluxetable*}

\end{document}